  \documentclass[twocolumn,prl,aps,showpacs]{revtex4}

 \usepackage[dvips]{graphicx}
 \usepackage[dvips]{graphics}
\usepackage{isolatin1}

\begin{document}

\title{Confined spin-waves reveal an assembly of nanosize domains in ferromagnetic La$_{1-x}$Ca$_x$MnO$_3$ (x=0.17, 0.2)}
\author{M. Hennion$^1$, F. Moussa$^1$, P. Lehouelleur$^1$, F. Wang$^{1,2}$, A. Ivanov$^3$, Y. M. Mukovskii$^4$ and D. Shulyatev$^4$}

\affiliation{$^1$ Laboratoire L\'eon Brillouin, CEA-CNRS, CE-Saclay, 91191 Gif sur Yvette, France.\\
$^2$ Institute of Physics Chinese Academy of Sciences, 100080 Beijing, China\\
$^3$ Institut Laue-Langevin, 156 X, 38042 Grenoble Cedex 9, France\\
$^4$ Moscow State Steel and Alloys Institute, Moscow 119991 Russia
}

\begin{abstract}
We report a study of spin-waves in ferromagnetic La$_{1-x}$Ca$_{x}$MnO$_3$, at concentrations x=0.17 and x=0.2 very close to the metallic transition (x=0.225). Below T$_C$, in the quasi-metallic state (T=150K), nearly q-independent energy levels are observed. They are characteristic of standing spin waves confined into finite-size ferromagnetic domains, defined only in ({\bf a, b}) plane for x=0.17, and in all directions for x=0.2. They allow an
 estimation of the domain's size, a few lattice spacings, and of the magnetic coupling constants inside the domains. These constants, anisotropic, are typical
 of an orbital-ordered state, allowing to characterize the domains as "hole-poor". The precursor state of the CMR metallic phase appears, therefore, as an assembly of small orbital-ordered domains.      
\end{abstract}

\maketitle
In metallic La$_{1-x}$Ca$_x$MnO$_3$, for x$\approx$0.3, the origin of the colossal magneto-resistance (CMR) which occurs at the metal-insulator transition is still a challenge for theories. The role played by an inhomogeneous ground state has been outlined by several theoretical\cite{Moreo} and experimental works\cite{Lynn,de Teresa}. In the semi-conducting parent LaMnO$_3$, the magnetic coupling is of super-exchange (SE) type, ferromagnetic (F) in the ({\bf a, b}) plane (J$_{a,b}$$>0$) and antiferromagnetic (AF) along {\bf c} (J$_c$$<$0), consistent with the orbital ordering which occurs at the Jahn-Teller transition T$_{JT}$. Doping with holes induces a canted AF state (CAF). There, neutron diffuse scattering indicates the existence of magnetic inhomogeneities attributed to hole-rich clusters embedded in a hole-poor matrix\cite{Hennion}, and a new spin-wave branch appears at small-q around F Bragg peaks. This new branch coexists with the spin-wave branch close to that of LaMnO$_3$ (SE type of coupling), attributed to the hole-poor matrix\cite{Biotteau}, which disappears at x=0.125 (F state). The ferromagnetic state which occurs in the range 0.125$\le$x$<$0.225 shows a quasi-metallic behavior below T$_C$ and an insulating one at lower temperature\cite{Okuda}. There, several works have also proposed an inhomogeneous picture using NMR\cite{Papavassiliou}, Mossbauer\cite{Chechersky} or magnetization measurements\cite{Markovitch}.

In the present paper, we report a determination of spin-waves in ferromagnets La$_{1-x}$Ca$_x$MnO$_3$ at x=0.17 and 0.2 very close to the "true" metallic state (x$\ge$0.225). Below T$_C$, in the  quasi-metallic state, q-independent energy levels of magnetic origin are observed in the large q-range. The q-dependence of their intensity, their evolution with temperature and magnetic applied field suggest that they can be ascribed to standing spin waves characteristic of a confinement into small domains. The shape and the size (a few lattice spacings) of the domains can be estimated as well as the magnetic coupling inside the domains. All the observations are consistent with a picture of hole-poor and orbital-ordered domains.

Inelastic neutron experiments have been carried out at the reactor Orph\'ee (Laboratoire L\'eon Brillouin) and at the Institut Laue-Langevin, on triple axis spectrometers, using fixed final wave vector of neutrons varying from 1.05 to 4.1$\AA^{-1}$, and appropriate filters.
The T$_C$ values are 175K and 180K for x=0.17 and
 x=0.2 respectively. Both samples have been characterized by resistivity measurements. In La$_{0.83}$Ca$_{0.17}$MnO$_3$, the high-temperature pseudo-cubic structure, with lattice parameter $a_0$=3.89$\AA$, becomes orthorhombic at T$_{JT}$=240K, whereas such an orthorhombicity is unobservable  for La$_{0.8}$Ca$_{0.2}$MnO$_3$. All {\bf Q} values ({\bf Q}={\bf q}+$\tau$ and q=$\zeta$) are given in reduced lattice units with pseudo-cubic indexation.
The two single crystals are twinned, so that each direction is superimposed on related symmetry directions.
Data are fitted by a lorentzian shape weighted by a Bose factor. This analysis yields the energy E(q), the damping $\Gamma$(q) with $\Gamma$(q)/E(q)$\approx$1/5 and the intensity.

 We first consider La$_{0.83}$Ca$_{0.17}$MnO$_3$. The spin wave modes obtained at T=150K, 100K and 10K along [001]+[010]+[001] directions are reported in Fig. 1, left panel. At 150K and 100K, one may distinguish two dispersive modes labelled (1) and (3). They are very close to previous measurements at x=0.125 reported in orthorhombic indexation\cite{Biotteau}. This comparison allows us to assign the curve (1), dispersing up to $\approx$18 meV, to [100] or [010] directions, which have been shown to be equivalent, and the curve (3), with a down-turn at q=0.25, to the [001] or {\bf c} direction. We mention that this down-turn, also observed at x=0.125 and in the CAF state, is surprising for a F coupling. It denotes an underlying AF coupling reminiscent of the orbital ordering of LaMnO$_3$. The very new feature is the gap-opening in the dispersive curve (1). Within this gap lies a nearly q-independent energy level, (2), at a mean value of 4.5 meV at 150K. 
 With decreasing temperature, all the energies increase. 
At 100K and below, a single dispersion curve is observed for q$<$0.125, larger gaps open at q=1/8, 1/4 and 3/8 and the level labelled (2) appears modulated with a maximum close to 3/8. Measurements of spin waves along [110] and [111] at 150K also reveal q-independent energy levels lying at nearly the same values in the two directions.

\begin{figure}[t]
\centerline{\includegraphics[width=9 cm]{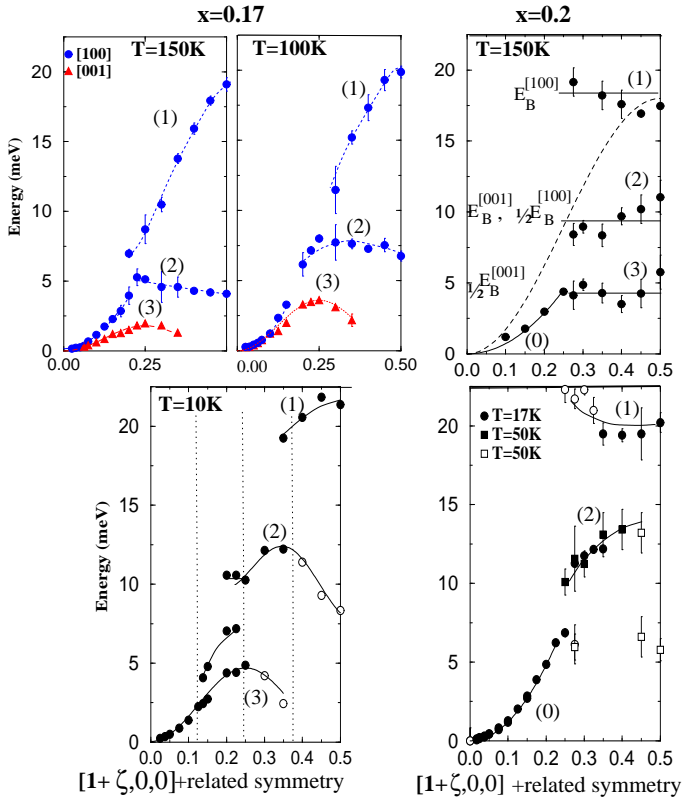}}
\caption{Magnetic excitations measured along [100]+[010]+[001] for x=0.17 (left panel) and x=0.2 (right panel) at several temperatures. Full (empty) circles correspond to modes with main (weak) intensity. In left panel, the solid and broken lines are guides to the eyes, and the vertical lines locate 1/8, 1/4 and 3/8 q values. In right-upper panel, the dashed line indicates the spin-wave dispersion of a virtual large-size domain (see the text).
}
\end{figure}

\begin{figure}[t]
\centerline{\includegraphics[width=8 cm]{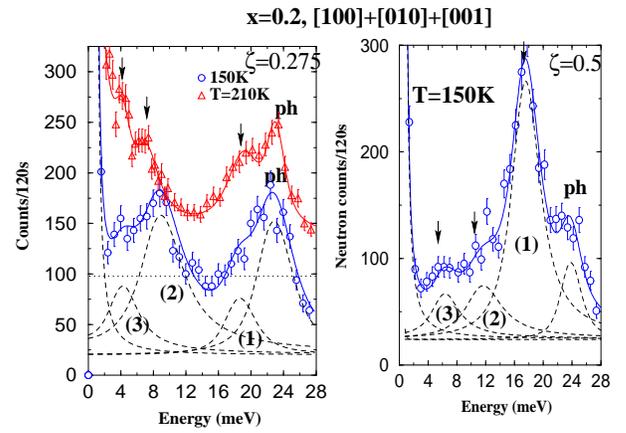}}
\caption{Examples of energy spectra along [100]+[010]+[001] at H=0, after application of a magnetic field H=2T. {\bf Left panel}: $\zeta$=0.275 at T=150K (circles) and 210K (triangles, shifted by 100 neutron counts). {\bf Right panel}: $\zeta$=0.5. The modes (1), (2) and (3) correspond to spin-wave branches indicated in Fig. 1, and "ph" is a phonon mode.
}
\end{figure}
 These observations call for the following comments. A q-independent energy level indicates localized excitations. The existence of a gap-opening in the branch (1) points out its interaction with level (2). This implies that level (2) belongs to the same macroscopic domain, namely to the ({\bf a, b}) plane, but not to the {\bf c} direction. The localized excitations or standing spin waves are confined within these planes, exhibiting a two-dimensionnal (2D) character. This is consistent with observations along [110] and [111], as indicated below. 
 Since the dispersionless level appears above q$_{cr}$$\approx$0.2,  
an approximate size $\xi$=$a_0$/q$_{cr}$$\approx$ 20$\AA$ may be provided for the domain. A more quantitative determination of the size can be done by considering the energy value of the q-independent level. Following the arguments developped below for the x=0.2 compound, the ratio, $\approx$1/4, between the energies of the level (2) and of the zone boundary, determines a size of 4 lattice spacings ($16 \AA$) along [100] (or [010]). In the insulating state and specially below 100K, this picture is no longer valid. We may explain the origin of gaps and the modulation of the level (2) by assuming an underlying periodicity of $4a_0$ in the ({\bf a, b}) plane. At the new zone boundaries of the super-cell, the dispersion is folded and new gaps occur. This description of the spin dynamics is likely related to observations in La$_{7/8}$Sr$_{1/8}$MnO$_3$, where static superstructures are observed\cite{Yamada}. 

The study of La$_{0.8}$Ca$_{0.2}$MnO$_3$ reveals a very interesting evolution with x. This sample has been first studied in zero field. Preliminary observations at 17K along [001]+[010]+[001] have been reported\cite{Biotteau}. Results are described with decreasing temperature. At 150K, unlike the x=0.17 case, a single dispersion branch, labelled (0), is observed in the small q range up to q$\approx$0.3, indicating that the [100] ({\bf a} and {\bf b}) and [001] ( {\bf c}) directions are now equivalent and the dispersion isotropic (see Fig 1-right upper panel). 
For q$\ge$ 0.3, the magnetic intensity is shared between several nearly q-independent levels defined by their mean values: a broad energy level, (1), at 18$\pm$1 meV, likely coupled to the lower optical phonon mode, a level (2) at 9$\pm$0.5 meV and an overdamped level, quasielastic. A magnetic field (H=2T) applied at 150K, suppresses the quasi-elastic intensity attributed to spin fluctuations, and reduces the energy linewidth of the dispersionless magnetic levels. When removing the field at 150K, the spin fluctuations remain considerably reduced, and the levels keep the characteristics observed in applied field. A low-energy mode (3) is now defined at a mean value 4.5 meV (see Fig 1) and the mode (1) appears separated from the phonon mode (24 meV), not reported in Fig 1. Corresponding energy spectra are displayed in Fig. 2, showing the q-dependence of the intensity. As q increases from q=0.275, to 0.5, the intensity of the modes (2) and (3) is transferred towards the energy mode (1) (18 meV), so that the intensity follows the q-dependence of a F dispersion curve. Very interestingly, these modes persist above T$_C$ (180K). This is shown by the spectrum at 210K in Fig. 2, left panel, where a slight decrease of the two energy levels (2) and (3) is observed whereas a quasielastic intensity (spin fluctuations) increases above T$_C$. In contrast, the small-q dispersive branch (0) renormalizes to zero at T$_C$. By lowering temperature in zero field, the lower-energy mode (3) nearly disappears and level (2) becomes more dispersed, which reduces the energy gaps (Fig. 1, lower-right panel).

Spin wave measurements along [110]+[101]+[011] and [111] directions are reported in Fig 3 at T=150K. At small q, the quadratic dispersion $\omega$=Dq$^2$, (0), defines an isotropic stiffness constant (D=30$\pm$1meV$\AA^2$ becoming 48meV$\AA^2$ at 10K). At larger q values, the magnetic intensity is transferred to nearly q-independent energy levels. Along [111], a level (2) is determined at 23$\pm$1 meV and a slightly dispersive one (1) around 46$\pm$2 meV. Fig 4 displays the balance in intensity between the modes (0) and (2) as q varies. Along [110]+[101]+[011], the entanglement between magnon and phonon branches (see dashed lines in Fig. 3-left panel), led us to use polarized neutrons (PN) which allow to identify magnons and phonons. Several levels have been determined, more or less dispersive. Dispersion could have been induced by the applied magnetic field (H=2T) which aligns macroscopic domains. Excitations at energy larger than 35 meV could not been detected because of a too weak intensity. By lowering temperature, the energy levels slightly increase.

\begin{figure}[t]
\centerline{\includegraphics[width=8 cm]{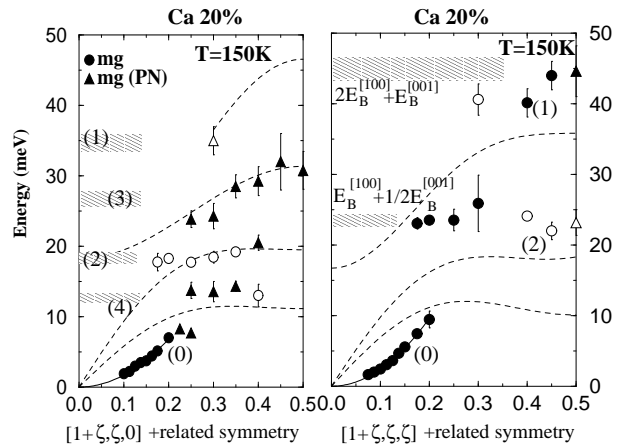}}
\caption{Magnetic (mg) excitations for q along [110]+[101]+[011] (left panel) and [111] (right panel) at T=150K, measured with unpolarised (circles) and half-polarized (triangles) neutrons. Full (empty) symbols refer to modes with main (weak) intensity. The hatched area correspond to calculated levels (see the text). Dashed lines are guide to the eyes for phonon branches measured at the same temperature. 
}
\end{figure}

\begin{figure}[t]
\centerline{\includegraphics[width=5 cm]{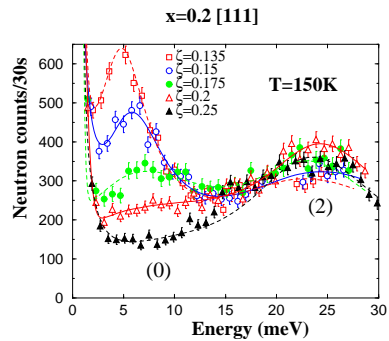}}
\caption{Energy spectra for several $\zeta$ along [111] at 150K, showing a dispersive mode (0) and a q-independent one (2).
}
\end{figure}

These overall observations can be explained as it follows.
Unlike La$_{0.83}$Ca$_{0.17}$MnO$_3$ where the confined spin waves have been assigned to the ({\bf a, b}) plane only (2D character), for x=0.2, the isotropy of the small-q dispersion suggests that all directions are concerned. The levels are attributed to standing spin-waves inside 3D clusters. Let us use the energy of the propagating wave for a 3D Heisenberg model (see dotted line in Fig. 1 upper-right panel):

 E(q$_x$,q$_y$,q$_z$)=8S\{J$_{a,b}$[1-0.5(cos(2$\pi$q$_x$)+cos(2$\pi$q$_y$))]+
J$_{c}$/2[1-cos(2$\pi$q$_z$)]\}, where q$_x$, q$_y$, q$_z$ are related to {\bf a, b, c} directions. This expression yields relations between J$_{a,b}$, J$_{c}$ and the zone boundary energies E$_B$: 
E$_B^{[100]}$=E$_B^{[010]}$=8SJ$_{a,b}$, E$_B^{[001]}$=8SJ$_{c}$, E$_B^{[110]}$=2E$_B^{[100]}$, E$_B^{[101]}$=E$_B^{[011]}$=E$_B^{[100]}$+E$_B^{[001]}$ and E$_B^{[111]}$=2E$_B^{[100]}$+E$_B^{[001]}$
These latter relations are also valid for finite-size clusters. According to predictions\cite{Hendriksen}, the cluster's size $\xi$ is related to the level with the lowest-energy value E$_L$. In the case of small clusters, it corresponds to a "half-wave" or a half-period of wave. For the confined wave of energy E$_B$, for which the nearest neighbour spins fluctuate in phase opposition, a half-period corresponds to one lattice spacing. Therefore, one expects that E$_B$ is a multiple of E$_L$. The number of lattice spacings defining the size may be provided by the ratio E$_B$/E$_L$. For a size of two lattice spacings along [100], one expects 2 levels, with values at E$_L$=E$_B$/2 and E$_B$, corresponding to confined half-wave and full-wave respectively. Along the other directions, if the cluster is isotropic or nearly cubic, one expects the same number of levels with the same ratio between the energies, the energy values being deduced from the above relations. We show now that the observations support this assumption.
In Fig. 1, right panel, level (1), E$_B^{[100]}$$\approx$18 meV, which defines the energy of the first neighbor coupling along [100], and level (2) at half this value, are assigned respectively to the full-wave and half-wave, standing along [100] (or [010] equivalent) confined in a cluster with 2 lattice spacings ($\xi$=8$\AA$). In the same way, level (2) at $\approx$9 meV and level (3) at half this value ($\approx$4.5 meV) are respectively assigned to the full and half wave along {\bf c} or [001]. Level (2) corresponds therefore to 1/2E$_B^{[100]}$ and E$_B^{[001]}$ values, superimposed because of twinning. From the E$_B^{[100]}$ and E$_B^{[001]}$ values, J$_{a,b}$ (1.12±0.1 meV) and J$_c$ (0.56±0.05 meV) are determined. Within the assumption of isotropic or nearly-cubic shape, two levels at E$_B$ and E$_B$/2 are also determined along the other symmetry directions. Due to the experimental uncertainty, they are shown by hatched area instead of narrow lines in Fig 3. In left panel, the sets (1), (2) and (3), (4), belong respectively to [110] and [101]+[011] directions, unequivalent, weighted by 1/3 and 2/3. The agreement with experimental values is good. It supports the proposed assignation of the levels and the assumption of isotropic shape for the clusters.
In the same way, for x=0.17, J$_{a,b}$ may be determined from E$_B^{[100]}$ (or E$_B^{[110]}$=2E$_B^{[100]}$) and J$_{c}$, from the difference between E$_B^{[111]}$ and E$_B^{[110]}$. These values are reported in Fig. 5, upper panel, with those obtained at x=0.2. 
The small value of J$_c$ is consistent with the 2D character found for the confined spin waves for this x=0.17 compound. 

\begin{figure}[t]
\centerline{\includegraphics[width=8 cm]{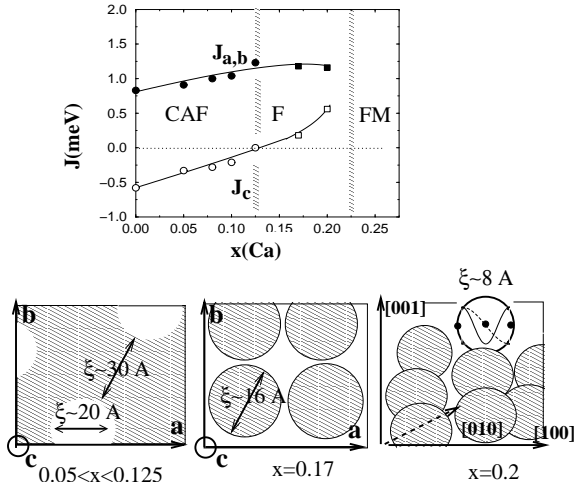}}
\caption{{\bf Upper panel}: Variation of the magnetic coupling values J$^{a,b}$ (full symbols) and J$^c$ (empty symbols) with x at 10K (circles) and 150K (squares). {\bf Lower panel}: Schematic drawing showing hole-rich (empty) and hole-poor (hatched) media. Increasing x from left to right, it shows 2D F hole-rich spin clusters, 2D F hole-poor spin clusters and isotropic 3D hole-poor spin clusters. A half (dashed line) and a full (solid line) standing waves are shown inside a cluster.
}
\end{figure}

The existence of clusters with a typical size indicates that the magnetic characteristics differ inside from outside the cluster. This implies a change of the charge density on the cluster's scale. The nature, "hole-poor", suggested by the anisotropy of the coupling constants determined above, is supported by a comparison with observations at lower doping. In Fig. 5 upper panel, the values J$_{a,b}$ and J$_c$, determined at 150K from the q-independent levels are reported with the J values obtained at 10K from the spin wave branch of SE type observed in the low doping and undoped compounds\cite{Biotteau}. Besides a small temperature effect (12\% in the 10K-150K range), the variation of J$_{a,b}$(x) and J$_c$(x) is monotonous across the CAF/F boundary (J$_c$=0 at x=0.125), showing a tendency to become isotropic at x=0.225 (metallic state). This variation implies a common origin for the coupling on the F and CAF sides, of SE type, characteristic of an orbital-ordered state with a roughly linear effect of the doping. This variation is close to theoretical predictions\cite{Feiner}. As in LaMnO$_3$ or in the CAF state, this SE coupling is attributed to "hole-poor" or without holes regions. Therefore, the cluster could consist of one Mn$^{3+}$ with its first Mn$^{3+}$ neighbourgs, the mobiles holes being confined at their boundaries. The small-q dispersion (0), which determines T$_C$, could be attributed to a coupling induced by mobile-holes through the clusters. These clusters differ from those found in the CAF state where the "hole-poor" region corresponds to the matrix and the clusters, described as "hole-rich" platelets\cite{Hennion}. The general evolution with x in direct space is schematically shown in Fig. 5. Finally, a comparison with other techniques\cite{Papavassiliou,Chechersky} allows to estimate the lifetime $\tau$ of these clusters for x=0.2, $10^{-6}s<\tau<10^{-9}s$, very large for the neutron probe. 

In conclusion, this study reports a quantitative description of the precursor state of the metallic phase which occurs for x$\ge$0.225. An evolution of the shape and of the size of clusters with x is proposed, thanks to the existence of confined spin waves which are observed for the first time in small F clusters. 
Very interestingly, these magnetic clusters have features (size, isotropy) very similar to those observed at larger x in CMR compounds around T$_C$\cite{Lynn}. The present observations could be therefore crucial for the understanding of the CMR properties.

The authors are very indebted to L. P. Regnault and J. Kulda for their help for experiments with field and polarized neutrons, and N. Shannon, T. Ziman, D. Khomskii and A. M. Ole\'s for helpful discussions.

\end{document}